\documentclass[9pt,sigconf, nonacm]{acmart}
\usepackage{multirow}
\usepackage{amsmath}
\usepackage{subcaption}
\usepackage{algorithm}
\usepackage{algpseudocode}
\usepackage{pgfplots}
\usepackage[table]{xcolor}
\usepackage{makecell}
\usepackage{pgfplots}
\pgfplotsset{compat=1.18}
\usepackage{xcolor}

\definecolor{bar1}{HTML}{6ABBEF} 
\definecolor{bar2}{HTML}{6ABBEF} 
\definecolor{bar3}{HTML}{A9D18E} 
\definecolor{bar4}{HTML}{9DC3E6} 

\pgfplotsset{compat=1.18}
\AtBeginDocument{%
  }

\begin{document}

\title{VerilogCL: A Contrastive Learning Framework for Robust LLM-Based Verilog Generation}

\author{Yan Tan, Tong Liu, Xiangchen Meng and Yangdi Lyu$^{\dagger}$}
\affiliation{%
  \institution{Microelectronics Thrust, Hong Kong University of Science and Technology (Guangzhou)}
  \country{}
}

\begin{abstract}

Large Language Models (LLMs) have recently achieved strong performance in software code generation. However, applying them to hardware description languages (HDLs), such as Verilog, remains challenging because high-quality training data are relatively scarce. In practice, LLM-generated Verilog often contains syntactic or structural errors that either cause compilation failures or produce functionally incorrect designs, which limit its reliability in hardware design workflows.

In this work, we propose VerilogCL, an integrated framework that enhances Verilog code generation by explicitly learning the boundary between correct and erroneous RTL through contrastive learning and proactive error screening. Our approach introduces minimal-error data augmentation, generating paired training samples of correct RTL and minimally perturbed erroneous RTL to teach the model to recognize fine-grained distinctions between correct and erroneous code. We then apply contrastive learning to learn a clearer validity boundary in the representation space, improving the separation between correct and erroneous RTL code. In addition, we introduce a proactive screening module that combines semantic embeddings with token-level uncertainty features to filter low-confidence candidates during generation. Experiments on public benchmarks, including VerilogEval and RTLLM, show that our 7B-parameter model outperforms the evaluated open-source, Verilog-specialized, and commercial baselines in both compilation success rate and functional correctness.

\end{abstract}


\maketitle

\section{Introduction}
\label{sec:introduction}
Large language models (LLMs), such as CodeLlama~\cite{roziere2023code} and its variants ~\cite{guo2024deepseek, bai2023qwen}, have achieved significant success in software development by enhancing productivity through automated code generation. This success is largely built on their training over extensive open-source codebases, primarily in popular languages like Python and C++~\cite{Mastropaolo_Pascarella_Guglielmi_Ciniselli_Scalabrino_Oliveto_Bavota_2023, Nijkamp_Hayashi_Xiong_Savarese_Zhou}. Building on these advancements, researchers are increasingly exploring the use of LLMs for generating hardware description languages (HDLs). Early results suggest that LLMs can produce basic HDL structures and simple modules, indicating their potential as assistants for front-end digital design~\cite{Benchmark_RTL_2022, Dehaerne_Verilog_2023, OpenLLM, VerilogEval, Xie_2023, 2024origen}. However, reliable RTL generation remains far more challenging than conventional software code generation~\cite{chen2021codex}.

A major challenge is the limited availability of high-quality HDL training data. Unlike software repositories, publicly available Verilog corpora are limited in scale, uneven in quality, and often lack the supervision needed for robust model adaptation. More importantly, Verilog is not merely a language for sequential computation; it is used to describe hardware structure, timing behavior, and parallel signal interactions. As a result, even small local deviations can either make the code syntactically invalid or introduce subtle functional bugs that are difficult to detect before simulation and synthesis. For instance, the improper mixing of non-blocking and blocking assignments within sequential logic, or the confusion between logical and bitwise operators, may still pass compilation, yet they often lead to functionally incorrect RTL during simulation or synthesis.

\begin{figure}[htbp]
    \centering
    \includegraphics[width=\linewidth]{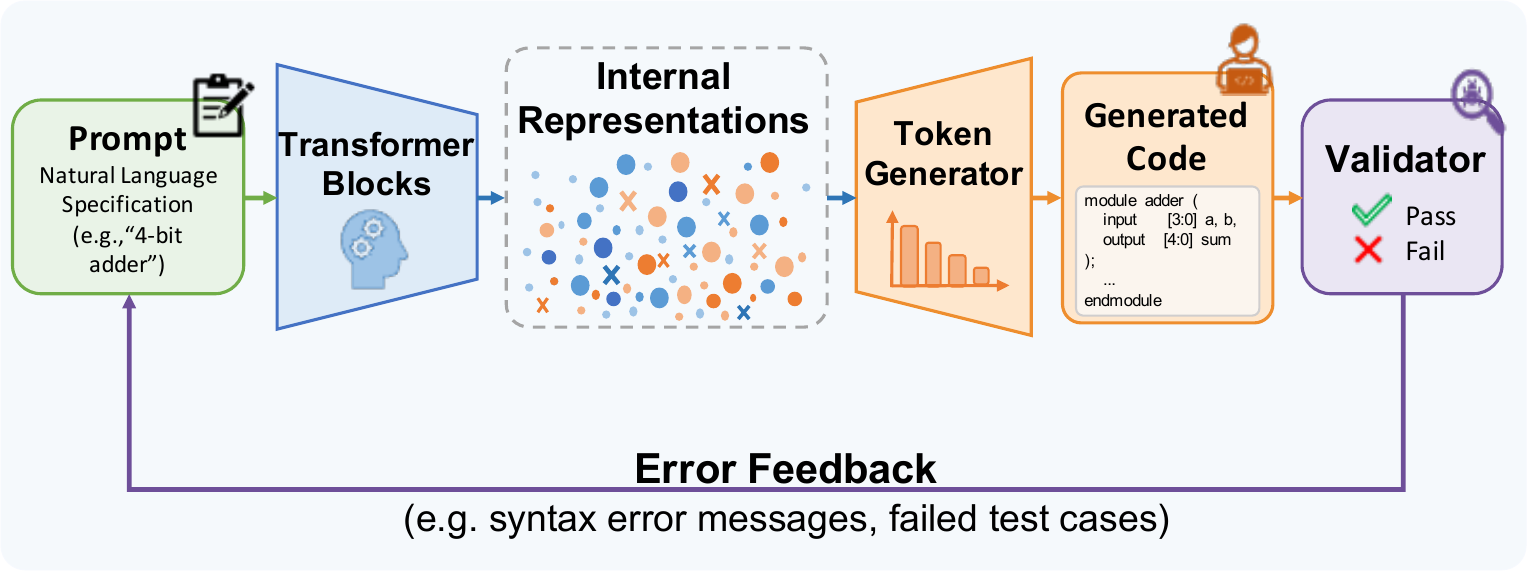}
    \caption{Existing LLM-Based Verilog Code Generation}
    \label{fig:existing}
\end{figure}


Hardware bugs are far more costly than software bugs because they can propagate into expensive EDA and manufacturing iterations. Accordingly, prior work has introduced error-feedback mechanisms that rely on external validation and compiler diagnostics to repair generated RTL~\cite{AutoChip,2024origen,RTLFixer,verilogcoder,wang2025hlsdebugger}. However, these methods remain largely post hoc: they repair errors after generation instead of improving the model’s representation of the boundary between correct and erroneous Verilog. Consequently, they still depend on an inefficient generate-compile-debug loop. We therefore argue that recurring errors stem not only from limited training data, but also from insufficient representation of fine-grained RTL validity boundaries.

To address this issue, we propose VerilogCL, a framework for robust Verilog generation that combines contrastive learning with inference-time candidate screening. Rather than relying solely on larger training corpora or iterative post-generation repair, VerilogCL is designed to improve the model's ability to distinguish correct RTL from minimally perturbed erroneous variants. We first construct minimal-error pairs, where each correct Verilog sample is matched with a counterpart containing a single well-defined perturbation. These pairs are used for contrastive training to pull correct samples closer together in representation space while pushing erroneous ones farther apart. In addition, we combine semantic representations with token-level uncertainty features to train a lightweight classifier for proactive screening during generation. Together, these components improve robustness by reducing error-prone continuations and increasing the likelihood of generating compilable and functionally correct RTL.

Our main contributions are summarized as follows.
\begin{itemize}
    \item We propose a novel data augmentation strategy that constructs paired samples consisting of correct RTL and minimally perturbed erroneous counterparts.
    \item We employ contrastive learning to refine the LLM's representation space, enforcing a clear boundary between correct and erroneous code representations. 
    \item We design a proactive screening mechanism based on hybrid features that combines internal semantic embeddings with statistical features derived from token-level uncertainty.
    \item Experimental results demonstrate that this framework significantly reduces errors in generated Verilog code and improves generation quality and reduces downstream invalid candidates.
\end{itemize}

\section{Background and Related Work}

\subsection{Verilog Code Generation with LLMs}
Recent advances in LLMs have shown strong potential for automating the generation of hardware description languages (HDLs), including Verilog. However, applying LLMs to RTL generation remains fundamentally challenging because it requires precise reasoning over structure, timing, and parallel behavior, while high-quality training data are still limited~\cite{chen2021codex,BetterV}. 

\subsubsection{\textbf{Data-Centric Approaches}} 
Data-centric approaches mainly improve generation by constructing hardware-specific datasets or applying augmentation. VerilogEval~\cite{VerilogEval} provides a curated supervised benchmark, while RTLCoder~\cite{Xie_2023} and OriGen~\cite{2024origen} further enrich training with aligned natural language–Verilog pairs or code-to-code augmentation. However, these methods largely treat Verilog as a text generation task and do not explicitly teach models to separate correct RTL from minimally perturbed erroneous RTL variants.

\subsubsection{\textbf{Correction-Centric Approaches}}

Correction-centric approaches improve correctness through post-generation repair or inference-time control. Works such as AutoChip~\cite{AutoChip}, RTLFixer~\cite{RTLFixer}, and OriGen~\cite{2024origen} use compiler feedback for iterative correction. Furthermore, some methods incorporate retrieval~\cite{BetterV,Chang_Wang_Ren_Wang_Liang_Han_Li_Li}, multi-stage debugging~\cite{wang2025hlsdebugger}, or agentic workflows~\cite{mage,vflow} to improve repair efficiency. Other studies move error handling into decoding, such as validity-aware temperature adaptation and AST-guided speculative decoding~\cite{DecoRTL,11133030}. However, these approaches mainly suppress or repair errors at the output level, rather than explicitly improving the model’s internal representation of the boundary between correct and erroneous RTL.

In summary, existing LLM-based Verilog generation methods mainly improve performance from two directions. Data-centric approaches enhance domain adaptation through curated supervision, augmentation, and fine-tuning, but they do not explicitly model the boundary between correct and erroneous RTL. This leaves an important gap: robust Verilog generation requires not only better supervision and stronger repair mechanisms, but also representation-level learning of syntactic and structural boundaries.

\subsection{Contrastive Learning}
Contrastive learning aims to bring representations of similar samples closer while pushing dissimilar ones apart, and has been widely used in representation learning~\cite{chen2021empirical,ye2019unsupervised,chen2020big}. In code-related tasks, it has proven effective for capturing structural and semantic relationships across programs ~\cite{jain2021contrastive,zhang2024code}. For example, UniXcoder~\cite{guo2022unixcoder} aligns representations across natural language, code, and structural signals, improving transferability and code understanding. These properties suggest that contrastive learning is suitable for modeling the fine-grained validity boundary required in Verilog generation.

Despite recent progress, existing contrastive learning studies mainly target general code semantics, program similarity, or multimodal alignment, rather than validity-aware representation learning for HDL generation. In Verilog generation, however, the critical challenge is ensuring that generated RTL strictly satisfies syntactic and structural constraints. To address this gap, we introduce contrastive learning in a validity-aware manner: instead of improving general code representations alone, we construct minimally perturbed correct/error pairs to explicitly teach the model to improve separability between correct and erroneous RTL. Unlike prior methods that rely on external correction or decoding-time constraints, VerilogCL reshapes the latent representation space itself, enabling the model to internalize Verilog’s syntactic and structural boundaries and reduce recurring errors at their source.

\section{Methodology}

\subsection{Overview}
\begin{figure*}
    \centering
    \includegraphics[width=1\linewidth]{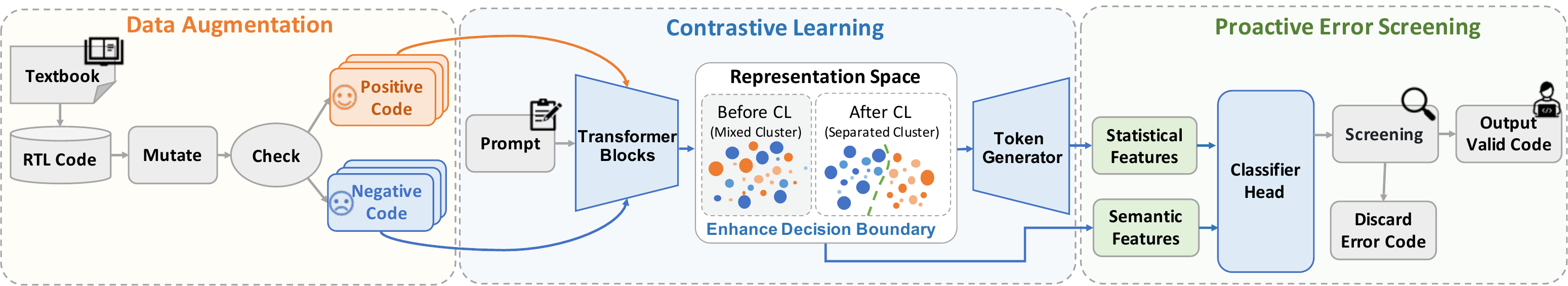}
    \caption{Overview of VerilogCL. The framework combines minimal-error augmentation, contrastive learning, and proactive error screening to shift part of error handling from post-hoc repair to generation-stage risk identification.}
    \label{fig:framework}
\end{figure*}

We propose VerilogCL, a validity-aware contrastive learning framework designed to improve the quality of LLM-generated Verilog, as illustrated in Figure~\ref{fig:framework}. The framework consists of three components. First, we construct minimal-error variants to augment the training data and expose the model to fine-grained RTL validity cues. Second, we apply contrastive learning to reshape the model’s representation space, making the boundary between correct and erroneous RTL  more explicit. Third, we combine hidden-state semantic features with uncertainty-based statistical features to train a classifier for inference-time proactive validity screening.


\subsection{Data Augmentation}

Unlike natural language tasks supported by large-scale corpora, publicly available Verilog datasets are relatively limited in scale and diversity, making it difficult for LLMs to learn fine-grained syntactic and structural constraints through simple memorization alone. To address this, we construct a minimal-error contrastive dataset that explicitly distinguishes correct RTL from closely related erroneous variants, enabling the model to learn a sharper boundary between correct and erroneous RTL. Our dataset covers two representative classes of RTL designs: combinational logic and sequential logic. The combinational subset includes Boolean operations, arithmetic circuits, multiplexers, and codec-related modules, while the sequential subset includes stateful elements such as D flip-flops, counters, memories, and finite state machines (FSMs). The detailed module categories are summarized in Table~\ref{tab:hardware_modules}.

\begin{table}[ht]
    \centering
    \renewcommand{\arraystretch}{1.2}
    \caption{RTL Module Categories Used in Data Augmentation}
    \vspace{-0.05in}
    \begin{tabular}{|c|p{2cm}|p{3.2cm}|}
    \hline
    \textbf{Category}       & \textbf{Module Name}                          & \textbf{Description} \\
    \hline
    \multirow{4}{*}{\textbf{Combinational}} 
        & Boolean\hspace{0.2in}Functions          & and\_gate, or\_gate, not\_gate, xor\_gate \\
        \cline{2-3}
        & Arithmetic       & half\_adder, full\_adder, comparator\\
        \cline{2-3}
        & Data Path                  & mux\\
        \cline{2-3}
        & Codecs                     & decoder, encoder \\
    \hline
    \multirow{4}{*}{\textbf{Sequential}} 
        & Storage                    & d\_flip\_flop\\
        \cline{2-3}
        & Counters                   & counter\\
        \cline{2-3}
        & Memory                     & ram, rom\\
        \cline{2-3}
        & FSMs & traffic\_light\_controller\\
    \hline
    \end{tabular}
    \label{tab:hardware_modules}
\end{table}

For each reference design, we construct contrastive samples consisting of one anchor, multiple positives, and multiple negatives. The anchor is a verified correct Verilog implementation. Positive samples are generated from the anchor using semantics-preserving transformations, such as variable renaming and equivalent logic restructuring, and are validated by equivalence checking. Negative samples are generated using script-based mutation rules derived from predefined Verilog error categories, including punctuation errors, keyword-related errors, operator misuse, declaration errors, and structural errors, as summarized in Table~\ref{tab:error_types}. The mutation rules were designed to reflect common failure patterns observed in preliminary generations of the base model and prior RTL-generation studies. Each negative sample introduces a single controlled perturbation, so that it remains close to the anchor while differing in RTL correctness.

\begin{table}[ht]
    \centering
    \caption{Verilog error categories used in minimal-error data augmentation.}
    \begin{tabular}{|p{2cm}|p{5.9cm}|}
    \hline
    \textbf{Error Type} & \textbf{Specific Errors} \\
    \hline
    \textbf{Punctuation errors} & Missing semicolons; missing commas in lists; extra commas at the end of lists; missing colons in width specifiers. \\
    \hline
    \textbf{Keyword errors} & Keyword typos; block delimiter mismatch; parentheses, brackets, or braces mismatch. \\
    \hline
    \textbf{Operator errors} & Assignment operator confusion; logical operator misuse; bitwise operator misuse; equality operator confusion. \\
    \hline
    \textbf{Declaration errors} & Missing signal declaration; incorrect signal type declaration; repeated signal declaration; missing or incorrect port direction declaration. \\
    \hline
    \textbf{Structural\hspace{0.05in}errors} & Assign statements outside module scope; multiple drivers for the same signal. \\
    \hline
    \end{tabular}
    \label{tab:error_types}
\end{table}

To ensure data quality, all mutated candidates are validated through compilation and simulation-based checking, and we retain only those that consistently produce compilation failures or functional mismatches. For each anchor design and each error category, we generate multiple mutated candidates, yielding a controlled set of erroneous RTL variants. Several mutation categories, such as operator misuse and declaration-related width inconsistencies, are associated not only with compilation errors but also with downstream functional mismatches. This construction provides structured supervision for contrastive learning by presenting the model with the differences between correct and erroneous RTL.

\subsection{Contrastive Learning Framework}

The goal of the contrastive learning framework is to sharpen the representation-space boundary between correct and erroneous RTL. By introducing Verilog-specific validity cues, the framework encourages the model to better distinguish valid RTL from erroneous variants and to capture the syntactic and structural constraints that are important for correct RTL generation, rather than relying primarily on memorization over limited-domain training data.

To explicitly separate correct and erroneous RTL in the latent space, we adopt a triplet margin loss. 
Given an input Verilog sequence \(x\), we first obtain the final-layer hidden states of the backbone LLM,
\[
H_x \in \mathbb{R}^{L \times D},
\]
where \(L\) denotes the sequence length and \(D\) denotes the hidden dimension. 
We then apply max pooling over the sequence dimension to obtain a fixed-length sequence representation:
\[
f(x) = \mathrm{MaxPool}(H_x) \in \mathbb{R}^{D}.
\]
Here, \(f(\cdot)\) serves as the embedding function used in contrastive learning. 
We use the same pooled semantic representation in both the contrastive learning stage and the downstream screening module, ensuring representation consistency across training and inference.

\begin{figure}[t]
    \centering
    \includegraphics[width=0.9\linewidth]{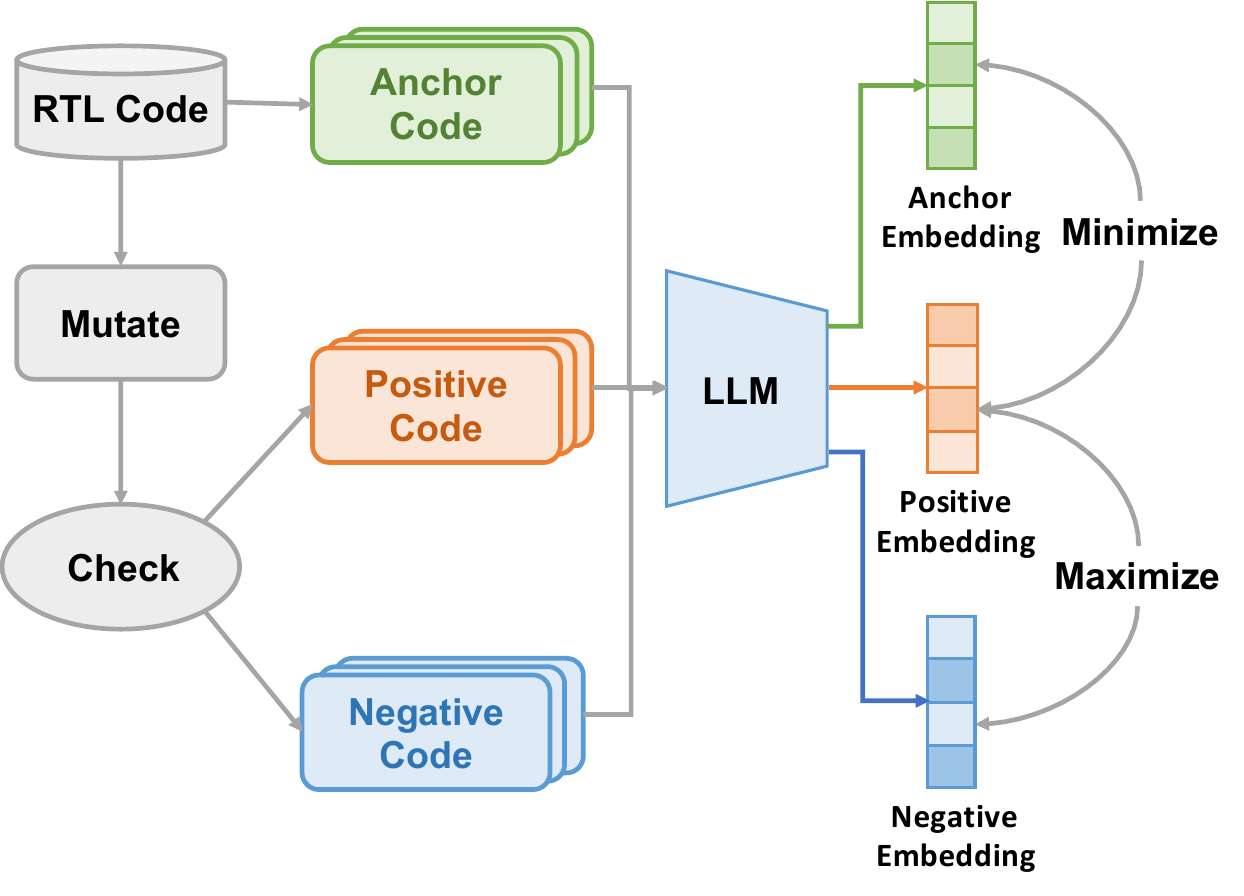}
    \caption{Triplet construction for minimal-error contrastive learning. A correct RTL sample serves as the anchor. Semantics-preserving transformations produce the positive sample, while a single injected error produces the negative sample.}
    \label{fig:contrast}
\end{figure}

Based on this representation, we define the triplet margin loss as
\[
L_{CL} = \max\!\bigl(0,\; \|f(a) - f(p)\|_2 - \|f(a) - f(n)\|_2 + m \bigr),
\]
where \(m\) is the margin hyperparameter, and \((a,p,n)\) denotes a triplet sample, as illustrated in Figure~\ref{fig:contrast}:
\begin{itemize}
    \item \(a\) (Anchor): a verified correct Verilog implementation;
    \item \(p\) (Positive): a semantics-preserving transformation of the anchor, such as variable renaming or equivalent logic restructuring;
    \item \(n\) (Negative): a minimally perturbed erroneous variant obtained by injecting a single well-defined error into the anchor, such as a mismatched keyword, an incorrect signal declaration, or an operator misuse.
\end{itemize}

Minimizing \(L_{CL}\) during training ensures that embeddings of correct RTL remain closer to the anchor than embeddings of erroneous RTL. This design is particularly suitable for Verilog generation, where many failures arise from small but validity-critical local deviations rather than large semantic changes. By constructing negatives with controlled minimal perturbations, the triplet objective explicitly teaches the model to capture fine-grained validity boundaries in RTL representation space. Moreover, several injected error types, such as operator misuse, declaration inconsistencies, and width-related errors, are closely related to downstream functional failures. Therefore, improving representation separability along these validity-related boundaries may also benefit correctness.

To efficiently adapt the pretrained LLM to the Verilog domain, we use Low-Rank Adaptation (LoRA). LoRA inserts trainable low-rank matrices into selected transformer layers, enabling targeted adaptation without updating the full parameter set. During training, triplets consisting of an anchor, a positive sample, and a negative sample are used to optimize the triplet margin loss. This parameter-efficient adaptation refines the representation space to better capture Verilog domain validity distinctions while reducing the risk of overfitting under limited-domain data.

\subsection{Proactive Error Screening}

Most existing LLM-based RTL generation frameworks detect errors only after a complete candidate has been generated and validated by external tools such as compilers or simulators. This post-hoc paradigm is inefficient because local decoding errors can propagate into full-sequence compilation failures or functionally invalid designs. To mitigate this issue, we introduce a proactive error screening module that identifies high-risk partial continuations directly from the model's internal signals before downstream validation.

The screening module is trained in a second stage after contrastive learning. We first fine-tune the LLM with the contrastive loss, then freeze the adapted model and extract semantic features for correct and erroneous RTL samples. These semantic features are combined with token-level uncertainty statistics to train a lightweight binary classifier. During inference, the adapted LLM generates RTL candidates, and the classifier is invoked at predefined statement boundaries to reject high-risk local continuations.

\subsubsection{\textbf{Hybrid Feature Representation}}

A key challenge in LLM-based code generation is that semantic similarity does not necessarily imply syntactic validity or functional correctness. In Verilog, two code snippets may express nearly identical design intent while differing only by a small syntactic or structural error, such as a missing delimiter, an incorrect declaration, or a width mismatch. As a result, hidden-state representations alone are often insufficient for distinguishing valid code from minimally erroneous variants.

To address this issue, we construct a hybrid representation that combines semantic features from hidden states with uncertainty-based statistical features derived from token-level output probabilities. Given the final hidden states \(H \in \mathbb{R}^{B \times L \times D}\), where \(B\) is the batch size, \(L\) is the sequence length, and \(D\) is the hidden dimension, we apply max-pooling over the sequence dimension for each sample to obtain a semantic representation. 
\[
v_{\text{sem}} = \mathrm{MaxPool}(H) \in \mathbb{R}^{D}.
\]

Semantic features capture global structural and contextual information from the generated RTL sequence.

We further extract statistical features $v_{\text{stat}}$ from the generation probabilities, including sequence-level uncertainty, local confidence fluctuations, and token-specific uncertainty patterns. These features summarize sequence-level uncertainty, local confidence fluctuations, and token-specific uncertainty patterns derived from the model's output scores. Table~\ref{tab:statistical-features} summarizes the statistical features used in this study. The final hybrid representation is formed by concatenation:
\[
v_{\text{hybrid}} = [v_{\text{sem}}; v_{\text{stat}}].
\]

\begin{table}[t]
\caption{Statistical uncertainty features included in (\(v_\text{stat}\)).}
\label{tab:statistical-features}
\centering
\resizebox{0.98\linewidth}{!}{
\begin{tabular}{|l|l|}
\hline
\textbf{Feature Type}    & \textbf{Feature Name}              \\ \hline
\multirow{6}{*}{Aggregate Uncertainty Features} 
        & avg\_nll                         \\ \cline{2-2} 
        & avg\_entropy                     \\ \cline{2-2} 
        & max\_nll                         \\ \cline{2-2} 
        & max\_entropy                     \\ \cline{2-2} 
        & std\_nll                         \\ \cline{2-2} 
        & low\_confidence\_token\_count   \\ \hline
\multirow{5}{*}{Spike Detection Features}       
        & spike\_num                       \\ \cline{2-2} 
        & mean\_spike\_value               \\ \cline{2-2} 
        & std\_spike\_value                \\ \cline{2-2} 
        & max\_spike\_value                \\ \cline{2-2} 
        & max\_spike\_token\_uncertainty   \\ \hline
\multirow{3}{*}{Token-Specific Features}        
        & punct\_mean\_nll                 \\ \cline{2-2} 
        & keyword\_mean\_nll               \\ \cline{2-2} 
        & last\_k\_mean\_nll               \\ \hline
\end{tabular}
}
\end{table}

\subsubsection{\textbf{Classifier Head}}

After contrastive training, we freeze the adapted LLM and extract hybrid features from the augmented samples to train the classifier head separately. The classifier head is designed to predict whether a given Verilog code snippet is correct or erroneous based on the hybrid feature representation. The classifier consists of a projection head followed by a linear classification layer with sigmoid activation. The projection head reduces the dimensionality of the concatenated hybrid vector \(v_\text{hybrid} = [v_\text{sem}; v_\text{stat}]\) to a compact latent representation, and the final layer outputs \(P(y=1)\), the predicted probability that the input snippet is valid. The classifier is trained with valid code as positive samples and minimal-error variants as negative samples. The training objective minimizes the binary cross-entropy loss:

\[
L_{Cls} = -\frac{1}{N} \sum_{i=1}^N \left[y_i \log(\hat{y}_i) + (1-y_i) \log(1-\hat{y}_i)\right],
\]

where \(y_i\) is the true label (1 for valid, 0 for erroneous), and \(\hat{y}_i\) is the predicted probability of validity.

\subsubsection{\textbf{Error Screening}}

During autoregressive decoding, the classifier is applied whenever the current partial sequence reaches a statement boundary. In our implementation, statement boundaries are defined by common Verilog delimiters such as ``;'', ``end'', ``endcase'', and ``endmodule''. When decoding reaches a predefined statement boundary, the classifier evaluates the hybrid feature representation \(v_\text{hybrid}\) and produces a validity score. If the score exceeds a preset threshold, decoding proceeds normally. Otherwise, the current partial continuation is marked as high risk, and the decoder rejects the current local continuation and resamples once from the current boundary under the same sampling configuration. This design enables the model to intervene at error-sensitive statement boundaries, thereby reducing the likelihood that local errors propagate into full-sequence compilation failures or functionally incorrect RTL implementations.

Proactive error screening therefore acts as an online risk-control mechanism that complements contrastive learning. Contrastive learning improves the separability between correct and erroneous RTL in latent space, while the screening module uses this improved representation together with uncertainty cues to reduce the chance that local decoding errors accumulate into invalid final outputs.

Algorithm~\ref{alg:verilogcl_generation} summarizes the inference-time proactive validity screening procedure. During generation, whenever decoding reaches a predefined statement boundary, semantic and statistical features are extracted from the current partial sequence and combined into a hybrid representation for classifier-based confidence estimation. This strategy reduces wasted downstream compilation on erroneous RTL candidates by steering inference toward higher-quality accepted samples.

\begin{algorithm}[t]
\caption{Proactive Validity Screening in VerilogCL}
\label{alg:verilogcl_generation}
\begin{algorithmic}[1]
\Require Input prompt $x$, Verilog generator $M$, screening classifier $C$, threshold $\tau$, maximum generation length $T$
\Ensure Generated Verilog code $y$

\State Initialize generated sequence $y \gets \emptyset$
\For{$t = 1$ to $T$}
    \Comment{\textcolor{blue}{Autoregressive decoding}}
    \State Decode one step conditioned on $(x, y)$, and obtain the updated sequence $y$, hidden states $H$, and output scores $S$
    
    \If{$y$ reaches a statement boundary}
    
        \Comment{\textcolor{blue}{Feature extraction}}
        \State Extract semantic feature via max-pooling:
        \State \quad $v_{\text{sem}} \gets \text{MaxPool}(H)$
        \State Extract statistical features from output scores:
        \State \quad $v_{\text{stat}} \gets \text{ComputeStats}(S)$
        \State Construct hybrid feature: $v_{\text{hybrid}} \gets [v_{\text{sem}}; v_{\text{stat}}]$
    
        \Comment{\textcolor{blue}{Proactive classifier screening}}
        \State Compute validity score:
        \State \quad $s \gets C(v_{\text{hybrid}})$
        \If{$s < \tau$}
            \State Reject the current local continuation and resample once from the current boundary
        \EndIf
    \EndIf

    \If{$y$ reaches EOS}
        \Comment{\textcolor{blue}{Return final output}}
        \State \Return $y$
    \EndIf
\EndFor
\State \Return $y$
\end{algorithmic}
\end{algorithm}

\section{Experiments}

\subsection{Experimental Setup}

We used DeepSeek-Coder-7B-Instruct-v1.5~\cite{guo2024deepseek} as the base model in all experiments.  To adapt it efficiently to the Verilog domain, we fine-tune it with LoRA in bfloat16 precision. LoRA adapters are applied to the attention projection layers with rank 32 and dropout 0.05. Training uses AdamW with a learning rate of $3 \times 10^{-5}$, batch size 8, and 10 epochs. At inference time, we use temperature 0.7 and top-p sampling with \(p=0.95\). All experiments are conducted on a single NVIDIA A30 GPU.


We construct the augmented dataset from verified reference implementations spanning 15 module categories (Table~\ref{tab:hardware_modules}). Negative samples are generated by rule-based error injection according to the categories in Table~\ref{tab:error_types}, with ten candidate variants per error type. To ensure data quality, all mutated candidates were subjected to compilation and simulation-based functional validation, and we retained only those variants that consistently produced either compilation failures or functional mismatches. Positive samples are generated by semantics-preserving transformations, such as variable renaming and equivalent logic restructuring, and are verified against the anchor designs using Yosys~\cite{yosys}. The final dataset contains approximately 3,000 contrastive training triplets.

To evaluate the performance of Verilog generation, we selected two representative benchmarks: VerilogEval~\cite{VerilogEval} and RTLLM~\cite{lu2024rtllm}. These datasets were specifically designed to evaluate the functional correctness and quality of Verilog code generated by language models. For evaluation, we strictly used the official benchmark testbenches, and all simulations were executed with Synopsys VCS.

\subsection{Contrastive Learning Evaluation}

\begin{figure}[htbp]
\vspace{-0.05in}
    \centering
    \begin{subfigure}{\linewidth}
        \centering
        \includegraphics[width=0.83\linewidth]{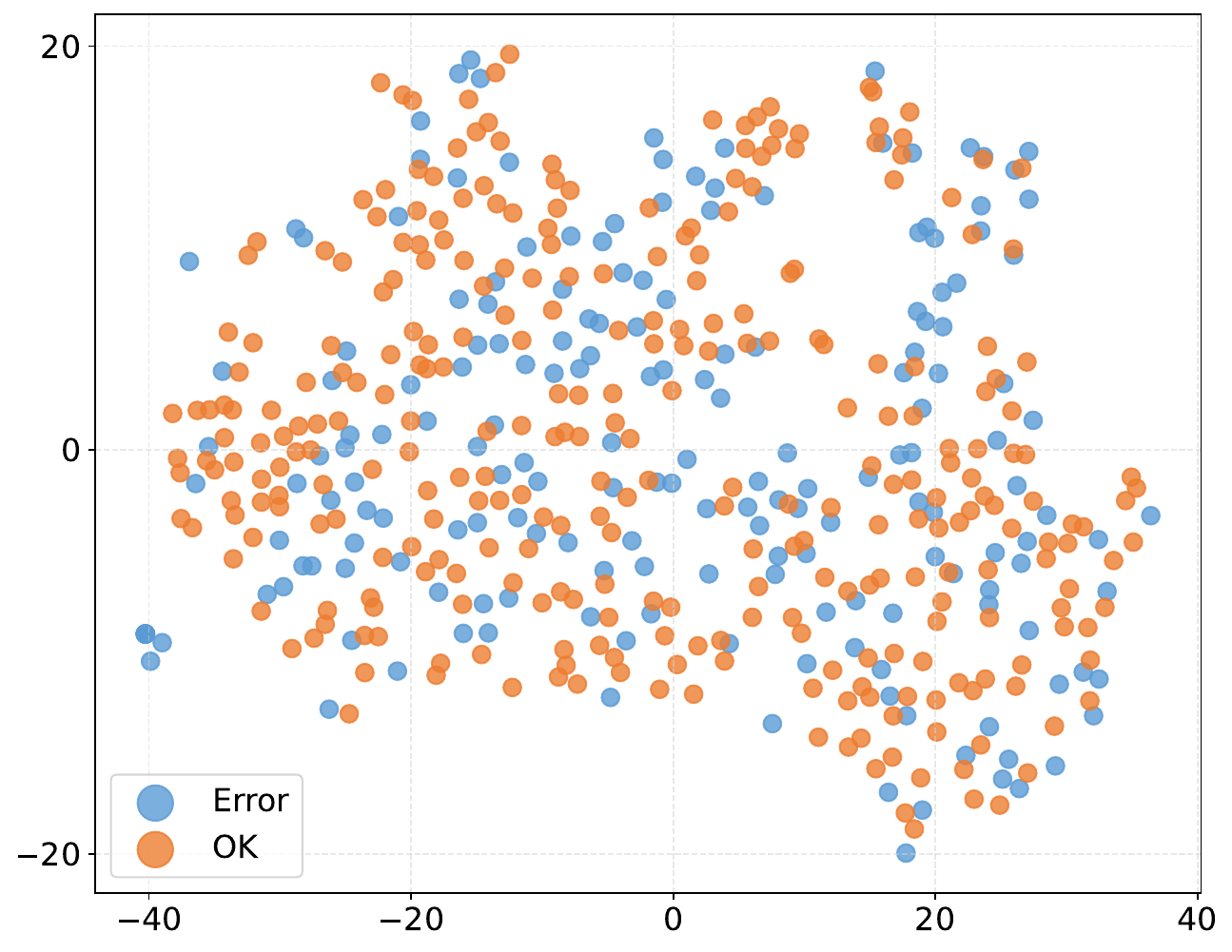}
        \caption{Before Contrastive Learning}
        \label{b10}
    \end{subfigure}
    \begin{subfigure}{\linewidth}
        \centering
        \includegraphics[width=0.8\linewidth]{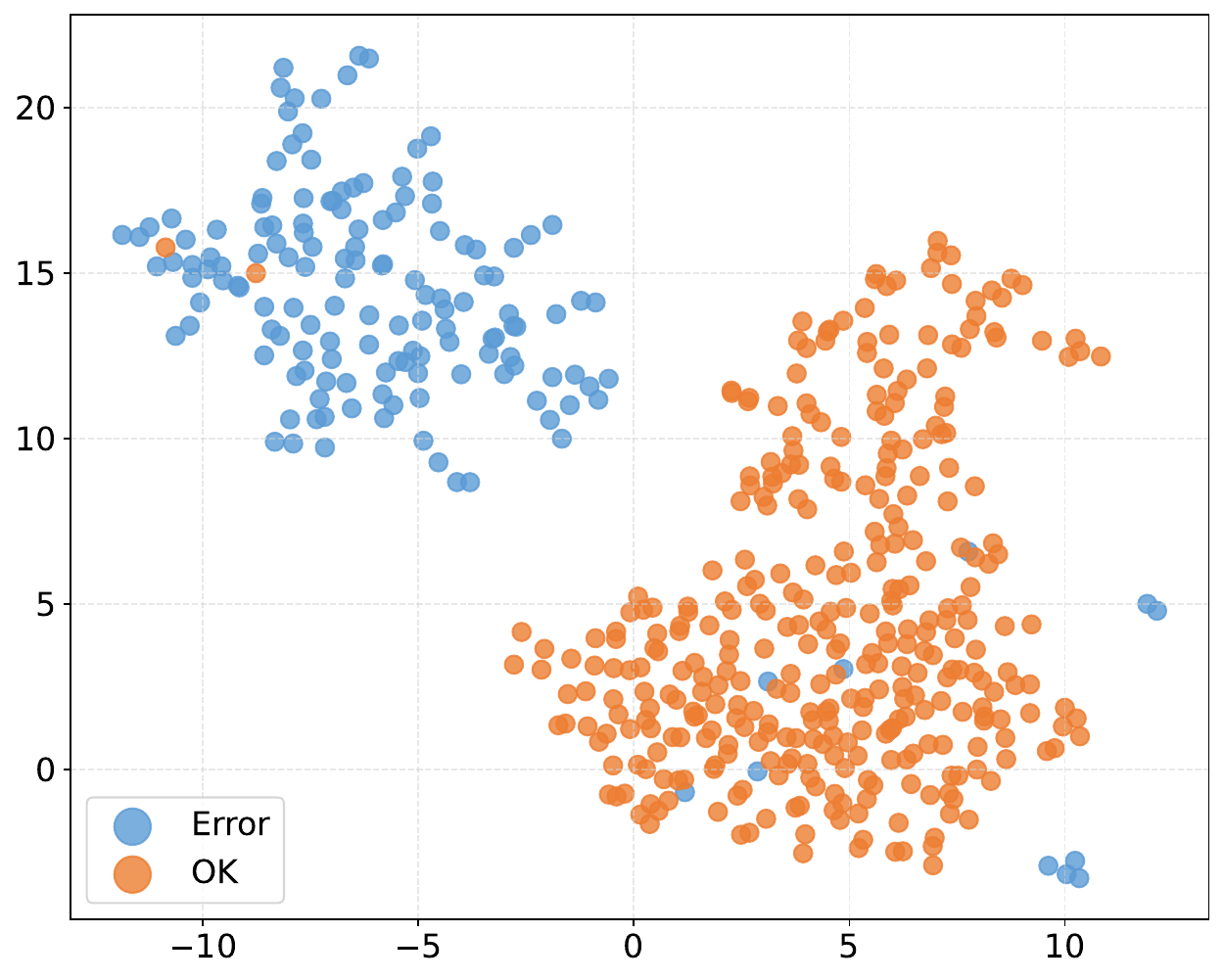}
        \caption{After Contrastive Learning}
        \label{b20}
    \end{subfigure}
    \caption{PCA visualization of semantic embeddings before and after contrastive learning.}
    \label{fig:pca}
\end{figure}

To evaluate the impact of contrastive learning on representation quality, we visualize the semantic embeddings of correct and erroneous RTL samples before and after contrastive training using PCA. As shown in Figure~\ref{fig:pca}, substantial overlap exists between correct (“OK”) and erroneous (“Error”) samples before contrastive learning. After contrastive training, the two groups become much more separable in the PCA projection, forming clearer clusters with reduced overlap.

This improved separability suggests that contrastive learning reshapes the semantic representation space to better distinguish differences between correct and erroneous RTL. In other words, code snippets that are semantically similar but differ in validity are more effectively separated after contrastive training. This improved representation quality provides a stronger foundation for the downstream screening classifier, which further incorporates uncertainty-based statistical features for more reliable proactive error screening during generation.

\subsection{Proactive Error Screening Results}
\label{subsec:screenresult}

We next evaluate the classifier used in the proactive screening module. The classifier is trained on the minimal-error paired dataset using hybrid features that combine semantic embeddings from LLM hidden states with uncertainty-based statistical features from token probabilities. The extracted features are split into 80\% training and 20\% validation sets. We train the classifier with a learning rate of 1e-4, batch size 8, dropout 0.1, and 100 epochs.

\begin{figure} [t]
    \centering
    \includegraphics[width=0.9\linewidth]{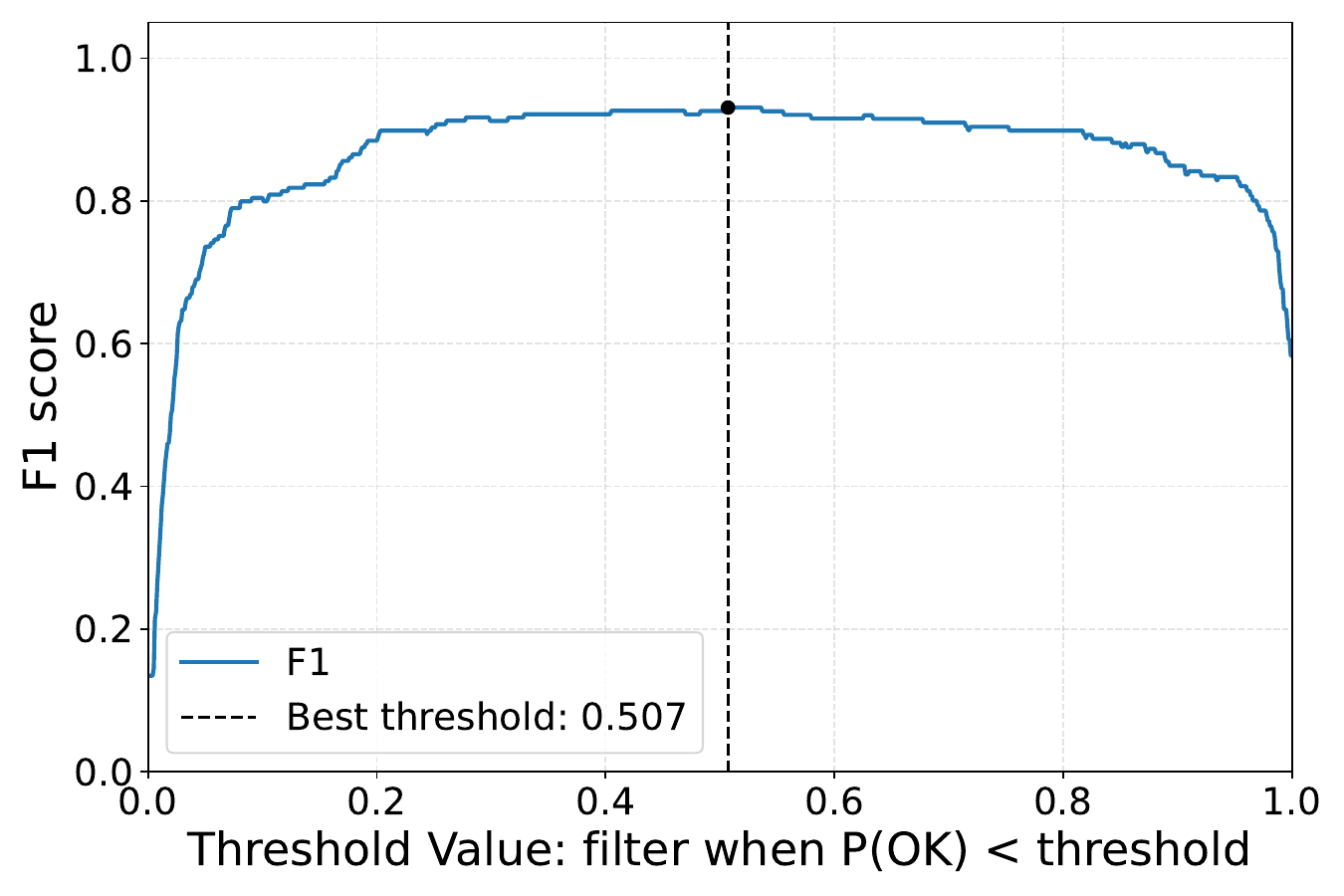}
    \caption{Validation-set F1 score under different screening thresholds on \(P(\mathrm{OK})\). The dashed line marks the optimal threshold \(\tau = 0.507\).}
    \label{fig:threshold}
\end{figure}

We first investigate the effect of different screening thresholds on classification performance on the validation set. As shown in Figure~\ref{fig:threshold}, the weighted F1 score on the validation set reaches its peak value of 92.58\% at a threshold of approximately 0.507. Performance remains relatively stable over a wide threshold range from 0.2 to 0.9, suggesting that the classifier is not overly sensitive to the exact threshold setting. For simplicity, we therefore set the screening threshold to \(\tau = 0.5\) in all subsequent experiments.

Table~\ref{tab:Classifier} summarizes the classification results on the validation set before and after contrastive training. Using the same classifier architecture and training protocol, we observe substantial improvements when the hybrid features are extracted from the contrastive-tuned model rather than the base model. Accuracy increases from 79.5\% to 92.5\%, and F1 improves from 79.96\% to 92.58\%. 

\begin{table}[htbp]
\centering
\caption{Classification performance before and after contrastive learning.}
\renewcommand{\arraystretch}{1.2}
\setlength{\tabcolsep}{10pt}
\begin{tabular}{>{\bfseries}lcc}
\toprule
\textbf{Metric} & \makecell[c]{\textbf{Before contrastive}\\ \textbf{learning}} & \makecell[c]{\textbf{After contrastive}\\ \textbf{learning}} \\
\midrule
Precision & 80.91\% & 92.80\% \\
Recall    & 79.50\% & 92.50\% \\
F1-score  & 79.96\% & 92.58\% \\
Accuracy  & 79.50\% & 92.50\% \\
\bottomrule
\end{tabular}
\label{tab:Classifier}
\end{table}

\begin{table*}[htbp]
\caption{Comparison of functional correctness on VerilogEval~\cite{VerilogEval} and RTLLM~\cite{lu2024rtllm}. The top scores ranked 1\textsuperscript{st}, 2\textsuperscript{nd}, and 3\textsuperscript{rd} in each column are highlighted in \colorbox[HTML]{FDE9E9}{Red}, \colorbox[HTML]{E0F7FA}{Blue}, and \colorbox[HTML]{E8F5E9}{Green}, respectively.}
\label{tab:comparison}
\centering
\small
\renewcommand{\arraystretch}{1.2}
\setlength{\tabcolsep}{2pt}
\begin{tabular}{p{2.8cm}|p{4.3cm}|ccc|ccc|c|c}
\toprule
\multirow{3}{*}{\textbf{Category}} & \multirow{3}{*}{\textbf{Model}} 
& \multicolumn{3}{c|}{\textbf{VerilogEval-human}} 
& \multicolumn{3}{c|}{\textbf{VerilogEval-machine}} 
& \textbf{RTLLM v1.1} 
& \textbf{RTLLM v2.0} \\

&  & \multicolumn{3}{c|}{\textbf{(\%)}} 
& \multicolumn{3}{c|}{\textbf{(\%)}} 
& \textbf{(\%)} 
& \textbf{(\%)} \\ \cline{3-10}

&  & \textbf{pass@1} & \textbf{pass@5} & \textbf{pass@10}
& \textbf{pass@1} & \textbf{pass@5} & \textbf{pass@10}
& \textbf{pass@5}
& \textbf{pass@5} \\ \hline

\multirow{5}{*}{\textbf{Verilog-Specific LLMs}} 
& VerilogEval~\cite{VerilogEval} & 28.8 & 45.9 & 52.3 & 46.2 & 67.3 & 73.7 & N/A & N/A \\
& CodeGen-6B MEV-LLM~\cite{nadimi2024multi} & 42.9 & 48.0 & 54.4 & 57.3 & 61.5 & 66.4 & N/A & N/A \\
& BetterV-CodeQwen~\cite{BetterV} & \cellcolor[HTML]{E8F5E9}46.1 & 53.7 & 58.2 & 68.1 & 79.4 & 84.5 & N/A & N/A \\
& RTLCoder~\cite{Xie_2023} & 41.6 & 50.1 & 53.4 & 61.2 & 76.5 & 81.8 & 41.7 & 53.6 \\
& OriGen~\cite{2024origen} & \cellcolor[HTML]{E0F7FA}54.4 & \cellcolor[HTML]{E0F7FA}60.1 & \cellcolor[HTML]{E0F7FA}64.2 & \cellcolor[HTML]{E0F7FA}74.1 & 82.4 & \cellcolor[HTML]{E8F5E9}85.7 & \cellcolor[HTML]{E0F7FA}65.5 & \cellcolor[HTML]{E0F7FA}58.8 \\ \hline

\multirow{3}{*}{\textbf{Commercial LLMs}} 
& Claude3-Sonnet~\cite{claude3_family} & \cellcolor[HTML]{E8F5E9}46.1 & \cellcolor[HTML]{E8F5E9}56.0 & \cellcolor[HTML]{E8F5E9}60.3 & 58.4 & 71.8 & 74.8 & 58.6 & 54.4 \\
& GPT-3.5~\cite{openai2023gpt35} & 35.6 & 48.8 & 52.6 & 49.4 & 72.7 & 77.6 & 44.8 & 36.2 \\
& GPT-4~\cite{openai2023gpt4} & 43.5 & 55.8 & 58.9 & 60.0 & 70.6 & 73.5 & \cellcolor[HTML]{E8F5E9}65.5 & \cellcolor[HTML]{E8F5E9}58.7 \\ \hline

\multirow{3}{*}{\textbf{Open Source LLMs}} 
& CodeLlama-7B-Instruct~\cite{roziere2023code} & 18.2 & 22.7 & 24.3 & 43.1 & 47.1 & 47.7 & 34.5 & 33.1 \\
& CodeQwen1.5-7B-Chat~\cite{bai2023qwen} & 22.4 & 41.1 & 46.2 & 45.1 & 70.2 & 77.6 & 37.9 & 36.4 \\
& DeepSeek-Coder-7B-Instruct-v1.5~\cite{guo2024deepseek} & 31.7 & 42.8 & 46.8 & 55.7 & 73.9 & 77.6 & 37.9 & 45.0 \\ \hline

\multirow{3}{*}{\textbf{Ours}} 
& DeepSeek-7B+Contrastive & 39.6 & 54.2 & 57.1 & \cellcolor[HTML]{E8F5E9}71.2 & \cellcolor[HTML]{E8F5E9}82.5 & 85.2 & \cellcolor[HTML]{E0F7FA}62.5 & 54.7 \\
& DeepSeek-7B+Classifier & 35.6 & 53.1 & 56.4 & 67.7 & \cellcolor[HTML]{E0F7FA}83.9 & \cellcolor[HTML]{E0F7FA}86.7 & 49.8 & 50.1 \\
& DeepSeek-7B+Contrastive+Classifier & \cellcolor[HTML]{FDE9E9}55.3 & \cellcolor[HTML]{FDE9E9}60.3 & \cellcolor[HTML]{FDE9E9}65.2 & \cellcolor[HTML]{FDE9E9}74.5 & \cellcolor[HTML]{FDE9E9}85.4 & \cellcolor[HTML]{FDE9E9}87.4 & \cellcolor[HTML]{FDE9E9}70.5 & \cellcolor[HTML]{FDE9E9}61.9 \\

\bottomrule
\end{tabular}
\end{table*}

These results indicate that contrastive learning improves the separability of the hybrid feature space and strengthens the classifier’s ability to distinguish valid from erroneous partial continuations. This improvement is important for inference-time screening, whose goal is not to certify final functional correctness, but to identify local high-risk continuations early and reduce error propagation during decoding. As a result, the screening module provides a stronger basis for improving downstream generation quality.

\subsection{Functional Correctness of Verilog Code}

Following VerilogEval~\cite{VerilogEval}, we use the metric \textit{pass@k} to evaluate the functional correctness of the generated Verilog code. \textit{pass@k} represents the expected probability of at least one answer passing the functional evaluation when randomly choosing $k$ answers from $n$ candidates:

\begin{equation}
    \text{pass@k} := \mathbb{E}_{\text{Problems}} \left[ 1 - \frac{\binom{n-c}{k}}{\binom{n}{k}} \right]
\end{equation}
where $n$ represents the total number of generated samples for a given problem, $c$ denotes the number of correct samples, and $k$ specifies the number of selected code samples. We set $n=10$ according to the original settings.


Table~\ref{tab:comparison} summarizes the functional correctness results on the evaluated benchmarks. Our full model, DeepSeek-Coder-7B-Instruct-v1.5 with contrastive learning and classifier-based screening, achieves the best overall performance among all compared methods. It obtains the highest pass@1 on VerilogEval-human and VerilogEval-machine, reaching 55.3\% and 74.5\%, respectively. On RTLLM v1.1 and RTLLM v2.0, where pass@5 is reported, our method also ranks first with 70.5\% and 61.9\%, respectively. Compared with the base DeepSeek-7B model, our full method improves pass@1 by 23.6 points on VerilogEval-human and 18.8 points on VerilogEval-machine, while improving pass@5 by 32.6 points on RTLLM v1.1 and 16.9 points on RTLLM v2.0. These results show that our method consistently improves functional correctness across different benchmarks.

We attribute these gains to two complementary factors. First, several mutation types used in contrastive learning, such as operator misuse, declaration inconsistencies, and width-related errors, are closely related to downstream functional failures in RTL. By explicitly separating these cases from correct code in the representation space, contrastive learning helps the model learn the boundaries between correct and erroneous RTL. Second, the proactive screening module steers decoding toward more reliable continuations by combining hidden-state semantics with token-level uncertainty features. This reduces the chance of following low-confidence continuations that often lead to erroneous RTL, thereby improving the overall quality of accepted outputs.

\subsection{Success Rate Evaluation}

To evaluate generation quality, we report \textit{success rate}, defined as the percentage of generated Verilog candidates that satisfy a given evaluation criterion. For each benchmark problem, we generate $N_{\text{total}} = 10$ candidates. The success rate is computed as
\begin{equation}
\mathrm{Success\ Rate} = \frac{N_{\text{success}}}{N_{\text{total}}} \times 100\%,
\end{equation}
where $N_{\text{success}}$ denotes the number of generated candidates that satisfy the target criterion, and $N_{\text{total}}$ denotes the total number of generated candidates. In this work, the criterion is either successful compilation or passing the official benchmark testbench. We compare VerilogCL against commercial LLMs, general open-source code models, and recent Verilog-oriented methods. Among the latter, OriGen~\cite{2024origen} and RTLCoder~\cite{Xie_2023} represent feedback-based RTL generation methods, while DecoRTL~\cite{DecoRTL} represents a validity-aware decoding approach.

\subsubsection{\textbf{Compilation Success Rate}}

Figure~\ref{fig:syntax} compares compilation success rates on the RTLLM v1.1 benchmark. VerilogCL achieves the highest compilation success rate among the evaluated methods, reaching 0.94 and outperforming RTLCoder (0.71), OriGen (0.78), DecoRTL (0.80), and all other baselines. Compared with the original DeepSeek-7B baseline (0.71), VerilogCL improves compilation success rate by 23 percentage points. These results indicate that our method is particularly effective at increasing the probability of generating compilable Verilog candidates.

\subsubsection{\textbf{Functional Success Rate}}

\begin{figure}[htbp]
    \centering

    \begin{subfigure}{\linewidth}
        \centering
        \begin{tikzpicture}
            \begin{axis}[
                ybar,
                bar shift=0pt,
                width=1.1\linewidth,
                height=4.5cm,
                ymin=0.2, ymax=1.0,
                ytick={0.20, 0.40, 0.60, 0.80, 1.00},
                yticklabel style={
                    /pgf/number format/fixed,
                    /pgf/number format/precision=2,
                    /pgf/number format/zerofill,
                    font=\normalsize
                },
                xticklabels={
                    Llama-7B,
                    Qwen1.5-7B,
                    DeepSeek-7B,
                    GPT-3.5,
                    GPT-4,
                    Claude3-Sonnet,
                    OriGen,
                    RTLCoder,
                    DecoRTL,
                    VerilogCL
                },
                xtick={1,2,3,4,5,6,7,8,9,10},
                x tick label style={
                    rotate=60,
                    anchor=east,
                    font=\normalsize,
                    align=right,
                    inner sep=1pt
                },
                bar width=13.5pt,
                enlarge x limits=0.08,
                nodes near coords,
                nodes near coords style={
                    font=\normalsize,
                    text=black,
                    yshift=1pt
                },
                grid style={line width=.1pt, draw=gray!30},
                ymajorgrids=true,
                axis line style={draw=gray!50},
                tick style={draw=none}
            ]
            \addplot[fill=bar3, draw=none] coordinates {(1, 0.35) (2, 0.54) (3, 0.71)};
            \addplot[fill=bar3, draw=none] coordinates {(4, 0.65) (5, 0.78) (6, 0.76)};
            \addplot[fill=bar3, draw=none] coordinates {(7, 0.78) (8, 0.71) (9, 0.80)};
            \addplot[fill=bar3, draw=none] coordinates {(10, 0.94)};
            \end{axis}
        \end{tikzpicture}
        \caption{Compilation success rate on RTLLM v1.1.}
        \label{fig:syntax}
    \end{subfigure}

    \vspace{0.5em}

    \begin{subfigure}{\linewidth}
        \centering
        \begin{tikzpicture}
            \begin{axis}[
                ybar,
                bar shift=0pt,
                width=1.1\linewidth,
                height=4.5cm,
                ymin=0.0, ymax=0.8,
                ytick={0.0, 0.20, 0.40, 0.60, 0.80},
                yticklabel style={
                    /pgf/number format/fixed,
                    /pgf/number format/precision=2,
                    /pgf/number format/zerofill,
                    font=\normalsize
                },
                xticklabels={
                    Llama-7B,
                    Qwen1.5-7B,
                    DeepSeek-7B,
                    GPT-3.5,
                    GPT-4,
                    Claude3-Sonnet,
                    OriGen,
                    RTLCoder,
                    DecoRTL,
                    VerilogCL
                },
                xtick={1,2,3,4,5,6,7,8,9,10},
                x tick label style={
                    rotate=60,
                    anchor=east,
                    font=\normalsize,
                    align=right,
                    inner sep=1pt
                },
                bar width=13.5pt,
                enlarge x limits=0.08,
                nodes near coords,
                nodes near coords style={
                    font=\normalsize,
                    text=black,
                    yshift=1pt
                },
                grid style={line width=.1pt, draw=gray!30},
                ymajorgrids=true,
                axis line style={draw=gray!50},
                tick style={draw=none}
            ]
            \addplot[fill=bar4, draw=none] coordinates {(1, 0.17) (2, 0.32) (3, 0.32)};
            \addplot[fill=bar4, draw=none] coordinates {(4, 0.34) (5, 0.52) (6, 0.47)};
            \addplot[fill=bar4, draw=none] coordinates {(7, 0.37) (8, 0.34) (9, 0.54)};
            \addplot[fill=bar4, draw=none] coordinates {(10, 0.57)};
            \end{axis}
        \end{tikzpicture}
        \caption{Functional success rate on RTLLM v1.1.}
        \label{fig:func}
    \end{subfigure}

    \caption{Success rate comparison on the RTLLM v1.1 benchmark.}
    \label{fig:success_rate}
\end{figure}
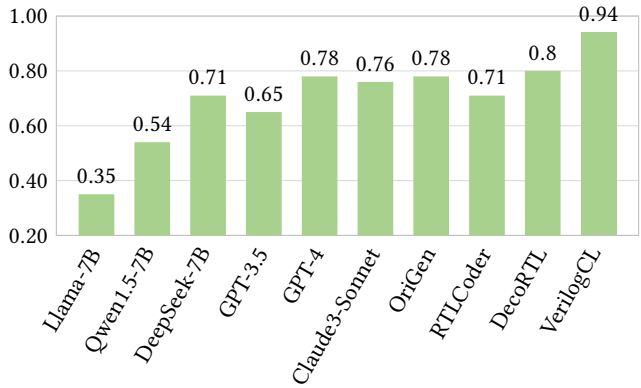
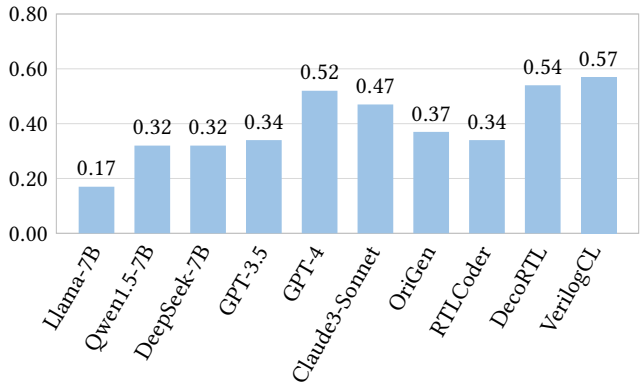

Figure~\ref{fig:func} compares functional success rates on the RTLLM v1.1 benchmark. Compared with compilation success rate, functional success rate is consistently lower across all evaluated methods, showing that successful compilation alone does not guarantee correct RTL behavior under simulation. Nevertheless, VerilogCL still achieves the best functional success rate, reaching 0.57 and outperforming DecoRTL (0.54), GPT-4 (0.52), Claude3-Sonnet (0.47), OriGen (0.37), and RTLCoder (0.34). Compared with the DeepSeek-7B baseline (0.32), VerilogCL improves functional success rate by 25 percentage points. This result suggests that the proposed framework improves not only syntactic reliability but also the likelihood of generating functionally correct RTL candidates.

\begin{table}[htbp]
    \centering
    \caption{Ablation study of compilation success rate and functional success rate on the RTLLM benchmarks. Base denotes DeepSeek-Coder-7B-Instruct-v1.5.}
    \label{tab:ablation}
    \renewcommand{\arraystretch}{1.1}
    \setlength{\tabcolsep}{3pt}
    \begin{tabular}{c|c|c|c}
        \toprule
        \textbf{Benchmark} & \textbf{Setting} & \textbf{Compilation} & \textbf{Functional} \\ \hline
        \multirow{4}{*}{RTLLM v1.1} 
        & Base        & 70.7\% & 32.1\% \\
        & Base+Contrastive & 78.8\% & 39.4\% \\
        & Base+Classifier  & 74.5\% & 36.2\% \\
        & Base+Both        & 94.0\% & 56.6\% \\ \hline
        \multirow{4}{*}{RTLLM v2.0} 
        & Base         & 69.8\% & 36.4\% \\
        & Base+Contrastive & 83.0\% & 39.4\% \\
        & Base+Classifier  & 78.8\% & 39.0\% \\
        & Base+Both        & 89.3\% & 49.8\% \\
        \bottomrule
    \end{tabular}
\end{table}

\subsubsection{\textbf{Ablation Analysis}}

To understand the contribution of each component, we perform an ablation study on RTLLM v1.1 and RTLLM v2.0 using compilation and functional success rates. As shown in Table~\ref{tab:ablation}, both contrastive learning and classifier-based screening improve over the base model, and their combination yields the best performance on both benchmarks. On RTLLM v1.1, the full method improves the baseline from 70.7\% / 32.1\% to 94.0\% / 56.6\%; on RTLLM v2.0, it improves from 69.8\% / 36.4\% to 89.3\% / 49.8\%, in compilation / functional success rate, respectively. These results indicate that the two components are complementary: contrastive learning improves feature quality and separability, while classifier-based screening reduces error-prone continuations during decoding.




\section{Conclusion and Future Work}

We present VerilogCL, a framework for improving the robustness of LLM-based Verilog generation through validity-aware contrastive learning and proactive error screening. By explicitly contrasting correct RTL with minimally perturbed erroneous variants, VerilogCL improves the model's ability to capture validity-relevant distinctions in representation space. Combined with inference-time screening based on semantic and uncertainty-aware features, this leads to fewer error-prone continuations during decoding. Experiments on VerilogEval and RTLLM show that the proposed framework substantially improves syntactic and functional correctness.

Future work will extend VerilogCL in two directions. First, we plan to apply the framework to more complex RTL generation settings, such as multi-module designs and longer-horizon generation tasks. Second, we will explore the integration of formal verification and design-aware objectives, including power, performance, and area (PPA), to support more reliable and implementation-aware RTL generation.


\bibliographystyle{IEEEtran}
\bibliography{references}

\end{document}